# SOCIAL MEDIA NETWORK ATTACKS AND THEIR PREVENTIVE MECHANISMS: A REVIEW

Emmanuel Etuh, Francis S. Bakpo and Eneh Agozie H

Department of Computer Science, Faculty of Physical Sciences,
University of Nigeria, Nsukka, Nigeria

***ABSTRACT***

*We live in a virtual world where actual lifestyles are replicated. The growing reliance on the use of social media networks worldwide has resulted in great concern for information security. One of the factors popularizing the social media platforms is how they connect people worldwide to interact, share content, and engage in mutual interactions of common interest that cut across geographical boundaries. Behind all these incredible gains are digital crime equivalence that threatens the physical socialization. Criminal minded elements and hackers are exploiting Social Media Platforms (SMP) for many nefarious activities to harm others. As detection tools are developed to control these crimes, hackers' tactics and techniques are constantly evolving. Hackers are constantly developing new attacking tools and hacking strategies to gain malicious access to systems and attack social media network thereby making it difficult for security administrators and organizations to develop and implement the proper policies and procedures necessary to prevent the hackers' attacks. The increase in cyber-attacks on the social media platforms calls for urgent and more intelligent security measures to enhance the effectiveness of social media platforms. This paper explores the mode and tactics of hackers' mode of attacks on social media and ways of preventing their activities against users to ensure secure social cyberspace and enhance virtual socialization. Social media platforms are briefly categorized, the various types of attacks are also highlighted with current state-of-the-art preventive mechanisms to overcome the attacks as proposed in research works, finally, social media intrusion detection mechanism is suggested as a second line of defence to combat cybercrime on social media networks*

## 1. INTRODUCTION

Social media network platforms provide mechanisms that enhance the effectiveness of virtual socialization in the global village. It is a medium that enable families, friends, and associates to interact and communicate seamlessly irrespective of their locations, distances, and platforms. Online Social Networks (OSNs) are the connection and communication platform that promotes the social interaction in the virtual space [1]. [2] identified 7 categories of social networks on the Internet to include: electronic mail services like Gmail, Yahoo mail, Microsoft outlook, Hotmail, etc; Instant messengers like WhatsApp, Twitter, Yahoo messenger, Instagram, Telegram, Snapchat, etc; Blogs platforms like Blogger, Tomblr, Wix, Linda etc; Social networking sites like Facebook, TikTok, Quora, LinkedIn, etc; Multimedia sharing systems like YouTube, Skype, Flikr, etc; Auction platforms like Jumia, Alibaba, Konga, OLX, etc; and Social search engines like Google, Yahoo, Safari, etc. All these platforms enable users to socialize and stay in touch with social reality in the virtual environment with varying functionalities. The pervasive nature of Information and Communication Technology (ICT) has greatly influenced every aspect of human activities; this has also influenced social media users to see the platform as a virtual home where they save their sensitive information on the database of these platforms.

The growing reliance on the use of social media networks worldwide has resulted in great concern for information security. One of the factors popularizing the social media platforms is how they connect people worldwide to interact, share content and engage in discussions of mutual interest that know no geographical boundaries. Behind all these incredible gains, most traditional crimes now have digital equivalence enabling criminal minded elements and hackers to exploit social media platforms for many nefarious activities to harm others. As security administrators and policy makers develop detection tools to control these crimes, hackers' tactics and techniques are also constantly evolving. Hackers are cybercriminals that specializes in virtual terrorism that endangers the legitimate users of Social Media Network Platform (SMNP) in particular and the entire virtual community in general [3] through various kinds of cyber-attacks.

These cybercrimes have significant negative impact on the social media platforms and the users in particular. Because of ease of accessibility, some of the social media users prefer to store their sensitive data on the network and when the account is hacked, this information could be used to swindle and defraud the user; also, the user's social contacts on the platform are at high risk of being defrauded by the hacker who could use their techniques by masquerading as the authorized user. High profile users like public and political leaders with private information that could tarnish their image if extracted can be used to threaten the user for ransom.

Different approaches have been used to prevent hackers' intrusion on the social media network platforms. The prevalent one is authentication using credentials like username and password, or PIN; biometric authentication like using face recognition technology, fingerprint, pattern matching, or voice recognition are all various forms of authentication. Other methods are Role Based Access Controls (RBAC), Extended RBAC, Temporal (RBAC), Risk-based access control [4]. These security methods have a lot of drawbacks like weak password which can easily be guessed by hackers using dictionary attack. In an attempt to enforce stronger password for authentication, social media users are forced to write their authentication credentials on papers which can also be stolen by hackers, these weaknesses have influenced many researchers to propose several security mechanisms to curtail the activities of these hackers. Some of these proposals include: biometric authentication, hybrid system for anomaly detection in social networks [5],Network Intrusion Detection System[6], [7], [8]etc.

All these methods are not suitable for data warehouse security. The commonly used network securities software like firewall and anti-viruses independently provide different services to network security but they can be bypassed by hackers. Hackers improvises new techniques of breaking into the social media platforms without being detected, the two close proposals on Data Warehouse Database Intrusion Detection System by [9] and [4] does not discourage resilient hackers. The limitations of all these proposed security approaches have been pointed out in Table 2.4 of the literature review. Hence, there is therefore a need for Intelligent Intrusion Detection Model (IIDM) that is efficient to disarm the hackers from carrying out their cybercrime activities against SMNP.

## 2. THEORETICAL BACKGROUND

Social media platforms have become an integral part of average network users in the virtual community. Billions of connected devices to the Internet operate on one social media platform or the other. According to report in [10], over 500million IoT devices were implemented globally in 2003, 12.5 billion in 2010, and 50 billion in 2020.There are about 3.5 billion people on social media with an estimated attacks that generate over $3 billion annually for cyber criminals [11]. Online social network platform like Facebook incorporate several functionalities like product and services advertisement and sales that makes it relevant to almost all internet users either cooperate or private. This has also increased cybercriminals' activity on the platform. According to a recent survey by Computer Emergency Response Team (CERT), the rate of cyber-attacks has been doubling every year [6]. Online social network is faced with threatening security challenges [2]. Facebook is the most popular social networking site. It was launched in February 2004 [12]With roughly 2.89 billion monthly active users as at the second quarter of 2021, Facebook is the biggest social network worldwide.

The Covid19 pandemic has been instrumental to the geometric shift to virtual socialization. The technological shift to cloud computing paradigm also has positively influenced the ubiquity of social media. This shift seems to have given hacker an edge to securely carryout their nefarious acts since humans are less involved. Cloud intrusion attacks are set of actions that attempt to violate the integrity, confidentiality or availability of cloud resources on cloud SMNP. The rising drop in processing and Internet accessibility cost is also increasing users' vulnerability to a wide variety of cyber threats and attacks.

Intrusion detection is meant to detect misuse or an unauthorized use of the computer systems by internal and external elements [7]. IDS are an effective security technology, which can detect, prevent and possibly react to the attack [13], [14] opined that artificial Intelligence plays a driving role in security services like intrusion detection.

### 2.1. Social media network

Social media network is a platform that creates virtual environment for social interactions among circle of friends and fans of like-minds. "Social media platforms are internet-based applications focused on broadcasting user-generated Content"[15] It deals with the sharing of information and multimedia content between users on similar platforms over electronic network especially the internet and cyberspace [16]. This platform has geometrically grown to become not only an effective communication tool for personal and social use, but also an essential channel for businesses and official communication channels. There are thousands of social media platforms being used today for different purposes, few of them that are most popular are highlighted below.

- **Facebook** is an online social media platform that provides several services like social networking of friends and fans, online advertising, voice calls, instant messaging, video calls, video sharing and viewing, online market place, virtual gifts among both young and the old, private and corporate bodies. It was launched on February 4, 2004, by Mark Zuckerberg. It had over 1.18 billion monthly active users as of August 2015 [16] and 2.85 billion active users in 20201 according to statistics by [17] with an engagement of over 4 billion views of videos everyday on the network. About 2.14 billion people can be reached via advertising on Facebook [17].
- **WhatsApp** is a cross-platform internet-based instant messaging application that allows smart phone users to exchange text, image, video and audio messages for free provided the device has Internet access. It was developed in 2009 by Brian Acton and Jan Koum. WhatsApp became the most popular messaging app with about 900 million active users as at September, 2015 [16].
- **MySpace** is a social networking website offering an interactive, user-submitted network of friends, personal profiles, blogs, groups, photos, music and videos. It was the biggest social media platform up till 2008 when it was overtaken by Facebook. It was cofounded by Chris DeWolfe and Tom Anderson

- **Twitter** is a social network platform that enables users to write and read short character messages called tweets. It revolves around the principle of followers who are equally users, who choose to follow another Twitter user and can thus view tweets sent by that user. Whereas unregistered users can read tweets, one must be registered to send tweets. It was founded in March, 2006 by Jack Dorsey [16].
- **Instagram** allows users to upload media that can be edited with filters and organized by hashtags and geographical tagging. Posts can be shared publicly or with pre-approved followers. Users can browse other users' content by tags and locations and view trending content. Users can like photos and follow other users to add their content to a personal feed. Instagram has 1.38 billion active users with 500 million daily active users of Instagram stories, 1.16 billion people can be reached through adverts on Instagram [17]
- **YouTube** is a video sharing service that allows users to watch videos posted by other users and upload videos of their own. With the ubiquitous use of smart phones this platform has become the first choice in personal broadcasting and video sharing. It was cofounded by Chad Hurley, Steve Chen, and Jawed Karim in February 2005. In November 2006, it was bought by Google and now operated by Google
- **LinkedIn** is a social media platform for professional networking. It is a social networking tool available to job seekers and professionals where users can invite other users and even non-users to connect. Inviters who get several rejections from invitees risk having their accounts restricted or closed. On this platform, users can get introduced to networks of contacts, new job and business opportunities, display products and services in their company profile pages, list job vacancies and search for potential candidates
- **Skype** is an IP telephony service provider that can be used to make free voice and video calls over the Internet to any Skype subscriber or to any other non-user at low calling rates. It is relatively simple to download and install the software, which works on most computers and phones. A dedicated Skype phone can be used or desktop computers, notebooks, tablets, mobile phones and other mobile devices fitted with a headset, speakers, microphones or USB phone. Skype also enables file transfers, texting, video chat and videoconferencing.
- **Viber** is a mobile application that allows phone calls and text messages to all other users, whether mobile or landline, for free. It is available over WiFi or 3G with sound quality much better than a regular call with mobile carrier charges applicable when used over a 3G network. Once the application is installed, calls can also be made to numbers that do not have Viber at low rates using ViberOut. Viber works on most android, iphone, blackberry, windows, mac, nokia and bada devices.
- **Tumblr** is a microblogging and social networking platform it's service allows users to post multimedia and other content to a short-form blog. Users can follow other users' blogs. Bloggers can also make their blogs private. For bloggers many of the website's features are accessed from a "dashboard" interface. It was founded by David Karp in 2007.
- **WeChat** is a Chinese multi-purpose instant messaging, social media and mobile payment app developed by Tencent. First released in 2011, it became the world's largest standalone mobile app in 2018, with over 1 billion monthly active users. It has been described as China's "app for everything" and a "super app" because of its wide range of functions. It provides text messaging, hold-to-talk voice messaging, broadcast (one-to-many) messaging, video conferencing, video games, sharing of photographs and videos and location sharing.
- **Reddit** This social media platform enables you to submit content and later vote for the content. The voting determines whether the content moves up or down, which is ultimately organized based on the areas of interest (known as subreddits). Number of active users per month: 100 million approximately.

- **Taringa** is one of the largest social networking platforms in Latin America and allows users to share their experiences, content and more. Number of active users: 75 million approximately.
- **Renren** is the largest social networking site in China and is literally a platform for everyone. It has been highly popular with the youth due to its similarity to Facebook, as it allows users to easily connect with others, quickly share thoughts and posts, and even update their moods. Number of active users per month: More than 30 million approximately

All the social media network platforms including the ones highlighted above can be categorized based on their support for the types of data they exchange or based on their aspect of support for social interactions.

Based on their support for the types of data they exchange, social media network platforms can be categorized into four main types:

i) Textual-based platform: used for text-related social communication for sending/receiving messages. A good examples are the messenger platforms.
ii) Visual-based platforms: used to for picture-related social interactions like sending and receiving images. A good example is flikr platform
iii) Audio-visual based platforms: used to for video-related social interactions like sending/receiving video data).A typical example is the YouTube,
iv) Hybrid platforms: this platform combines the functionalities of more than one of the textual, visual, and audio-visual platforms. A typical example is the Facebook

Social media platforms can also be categorized based on their aspect of support for social interactions. These categories are summarized in Table 2.1 below.

Table 2.1. Summary of Categories of Social Media Platform

| Category | Usage | Examples |
|---|---|---|
| Electronic Mail Service Platforms | The very first social media platform that gave birth to the Internet. Used to send and receive electronic mails from friends and associates | Gmail, Yahoo mail, Microsoft outlook, hotmail, etc |
| Social networking websites | Mostly used to connect friends, family, brands, and to reach out to target audience. | Facebook,Twitter, Whatsapp, Instagram, |
| Discussion Platforms | Mostly designed for research, and focused discussion with people with common interest | Reddit, Quora, Nairaland, etc |
| Blogging platforms | Mostly used for news, writing of articles, and personal messages to targeted audience | Tumblr, Medium, Blogger, Wix, Linda |
| Instant Messenger Platforms | Used for real-time textual conversation with instant sending and receiving functionalities | Whatsapp, Twitter, Yahoo messenger, Instagram, Snapchat, |
| Multimedia Platforms | For sharing of videos with both subscribed and visitor social media user of this platform | YouTube, Skype Flikr, TikTok, etc |
| Auction Systems | For sales of goods and services | Jumia, Alibaba, Konga, OLX, |
| Cooperate Social Platforms | Most companies are now incorporating social flavor like blog to their virtual platform to enable them to reach their target customers with their products and services | Microsoft news, Yahoo news, BBC news, Safari etc |
| Educational /Professional Platforms | For sharing knowledge through, chart, uploading of resources, live meetings, and connecting with professionals in the chosen profession of the social | Elsevier, Academia, Slack. DOAJ, Google Meet, Zoom, LinkedIn |
| | media user | etc |
| Gaming platforms | Hosts gamming applications where social media users play games either for entertainment, fun, or betting | Nairabet, Naijabet, Kingsbet, etc |
| Search Engines | It is the major tool that enhances virtual socialization. It enable social media user to easily locate the object of search | Google, Yahoo, Microsoft edge, Firefox, etc |

## 2.2. Social media network platforms attacks

There are several attacks launched against social media network platforms. It is important to know them because a more thorough understanding of these types of attacks equips social media user an armament of defensive measures and knowledge to lessen the likelihood of being exploited [3]. In August 6, 2009, Twitter, Facebook, LiveJournal, Google's Blogger, and YouTube were attacked by a distributed denial-of-service (DDoS) attack, in October, 2021, similar service disruption was encountered. [3] identified seven deadliest attacks on social media network platforms. These attacks are highlighted as follows

**1.** **Social networking infrastructure attacks**: here the attacker launches the attack on the platform that provides the social service with the view to disconnecting users from accessing the services provided by the platform. The major attack used against social networking infrastructure that directly affects the users is DDoS.

**2.** **Malware attacks:** in this type of attack, the hacker develop a malicious software with the intention of gaining control and utilizing the user's device to perform some malicious activities like launching DoS attack, keystrokes logging, theft of credential, credit-card number or bank details, etc. The mode of infecting user's device on social media is usually through links or images sent to the user's inbox knowing that the user will likely open since it comes from a connected social contact [18]. Once a user is infected, the hacker uses the compromised social media account to spread the worm by delivering a message to other users who are friends with the infected user containing a luring link to a third-party Web site, where they are then prompted to perform an action like "register to view full image", "update you Adobe Flash player to have a better view", etc. Once the action is performed, the worm will automatically infect the devices of all the connected friends that followed the link to the third party site. Common Malware Categories are Crimeware, Spyware, Adware, Browser Hijackers, Downloader, Toolbars, and Dialers [18]. Hackers leverage on the openness of social networking sites where users generate their contents; the large number of users; and the trust that is implied where users assume all friends are to be trusted to launch attacks to connected billions of users. The most effective method is by using Cross-Site Scripting XSS to implement their malicious codes on social networking site.

**3.** Phishing attacks: as the name imply is a hacking technique where the hacker lure a user using "bate" that is most appealing to the user with the intention of trapping the user. In most cases, the user is coarse into divulging sensitive information that will in turn be used to attack the user.

**4.** Evil twin attacks: in this type of attack, the hacker uses the target's profile to create account to mimic the authentic user. This attack can also be called cyber-impersonation. The new account is then used to send friend's request to the contacts on the social media platform just to enable the attacker to enjoy the privileges of friends and gain access to the users on the platform.

**5.** Identity theft: in this type of attack, the user's credential is stolen and used to securely gain access to the user's social media platform. Once the attacker successfully gains access, they launch their pre-conceived attacks while impersonating the authentic user.

**6.** Cyber-bullying: a way of threatening or intimidating a social media user either through messages or by posting objectionable content on the social media network to harass or intimidate the targeted user.

**7.** Physical threats: in this attack, the hacker launches physical attack against selected user. It could be in form of bypassing physical security of the platform through threatening the user to remove the device security.

All these threats can use any of the following attacking mode to carryout the threats.

1. Denial of Service (DoS) where an attacker tries to prevent legitimate users from using a service
2. Probe attack where an attacker tries to find information about the target host through ways such as scanning victims to get information about available services and the operating system
3. User to Root (U2R) where unauthorized access to local super-user privileges are being granted
4. Remote to Local (R2L) where unauthorized access from a remote machine through approaches such as guessing password to obtain a local account on the victim host
5. Advanced Persistent Threat (APT) is a targeted attack against a high-value asset or physical system where attackers often leverage stolen user credentials or zero-day exploits to avoid triggering alerts

### 2.3. Security measures for social media network

The “juicy prospect” of social media network platforms has made hackers to constantly device techniques to intrude and usurp users. They have two fold targets which are the social media users and the SMNP which they break into and control for their selfish gain [19]. On the users end, the hackers’ activities make them susceptible to threats which include identity theft, evil twin, password resetting, sim cloning, brute force, fake links, phishing, information leakage, celebrity spoofing, fake account, impersonation, etc [20] [21] [22]. They also use code injection through malicious SQL script to disrupt the network. The existing security mechanism for DW include Role Based Access Controls (RBAC), Extended RBAC, Temporal RBAC (TRBAC), Risk-based access control[4] which all has to do with authentication using username and password. [9] were the first to propose Database Intrusion Detection Systems (DIDS) for DW. [4] improved it by incorporating second level authentication, instant messenger like Whatsapp also uses two-steps verification where the user is asked to supply a PIN at intervals to prevent hackers from the network any time there is a detected anomaly, but hackers still use social engineering to trick the users thereby compromising the account which go undetected by role-based access controlled systems.

To hack into a social media network, there are basic steps taken by attackers to perform their operations. These steps are:

i. Target selection: the attacker determine the victim of the social media users to attack
ii. Attack selection: The attacker determine the type of attack to launch against the target, e.g infrastructure, malware, phishing, evil-twin, identity theft, cyber-bullying, physical threat, celebrity spoofing etc.
iii. Strategy formation: the strategy is dependent on the type of attack to use. If the selected attack is to launch DDoS attack, then the attacker will have to recruit accomplices (botnet) to use in launching the attack either through Calls, E-mail, Post to user groups, or Creation of a Web page where users are redirected to for infection.
iv. Army training: Furnish the accomplices with package containing the attack, time, date, and instructions on how to perform the attack.

v. Launching attack: here the attacker will launch the attack, wait and watch the execution of the attack.

To overcome the various attacks highlighted on the social media network platform, the user should:

i. Ensure an up-to-date antivirus software on the device as a primary line of defence
ii. Not open e-mails from people you don't know.
iii. Not click on unknown links
iv. Not visit unfamiliar sites
v. Disable JavaScript.
vi. Maintain and ensure regular update on software patches.
vii. Implement browser security policies, such as blocking pop-ups and limiting the number of connections.
viii. Implement platform privacy security policies, such as "who can see my personal info", "who can post on my wall", status, etc
ix. Implement IDS/IPS as second line of defence against attackers.

## 2.4. Intrusion detection mechanisms

Intrusion detection system (IDS) is a device or software application that monitors a system, network or application for malicious activity or policy violations with the aim of detect them. Two major IDS emphasized in literatures are Host-based and Network-based. These IDSs are not suitable for intrusion detection in application related intrusion attacks. This gave rise to the development of application specific IDS which is application based. Fig 2.1 shows the three types of intrusion detection systems presently available for information security.

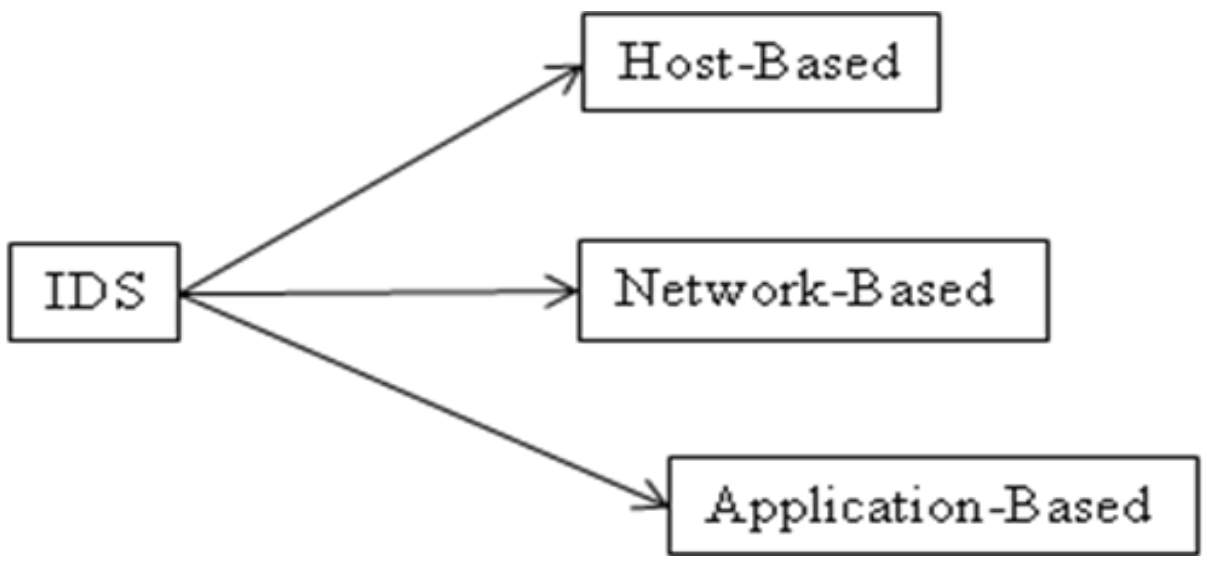

Fig 2.1. Types of Intrusion Detection System

Researchers have proposed various methods to detect abnormal operations in a system. Generally, IDS comprises of four main components: Traffic Collector, Analysis Engine, Signature Database, Management and Reporting Interface [7]. Network-based, Host-based, or Application-based intrusion detection systems use either signature mechanism to detect intrusion or anomaly approach. The signature approach uses rules for decision making on classifying intrusion based known profile of intrusion. Anomaly on the other hand classifies an operation as intrusion based on a deviation from the known normal operation of a given system.

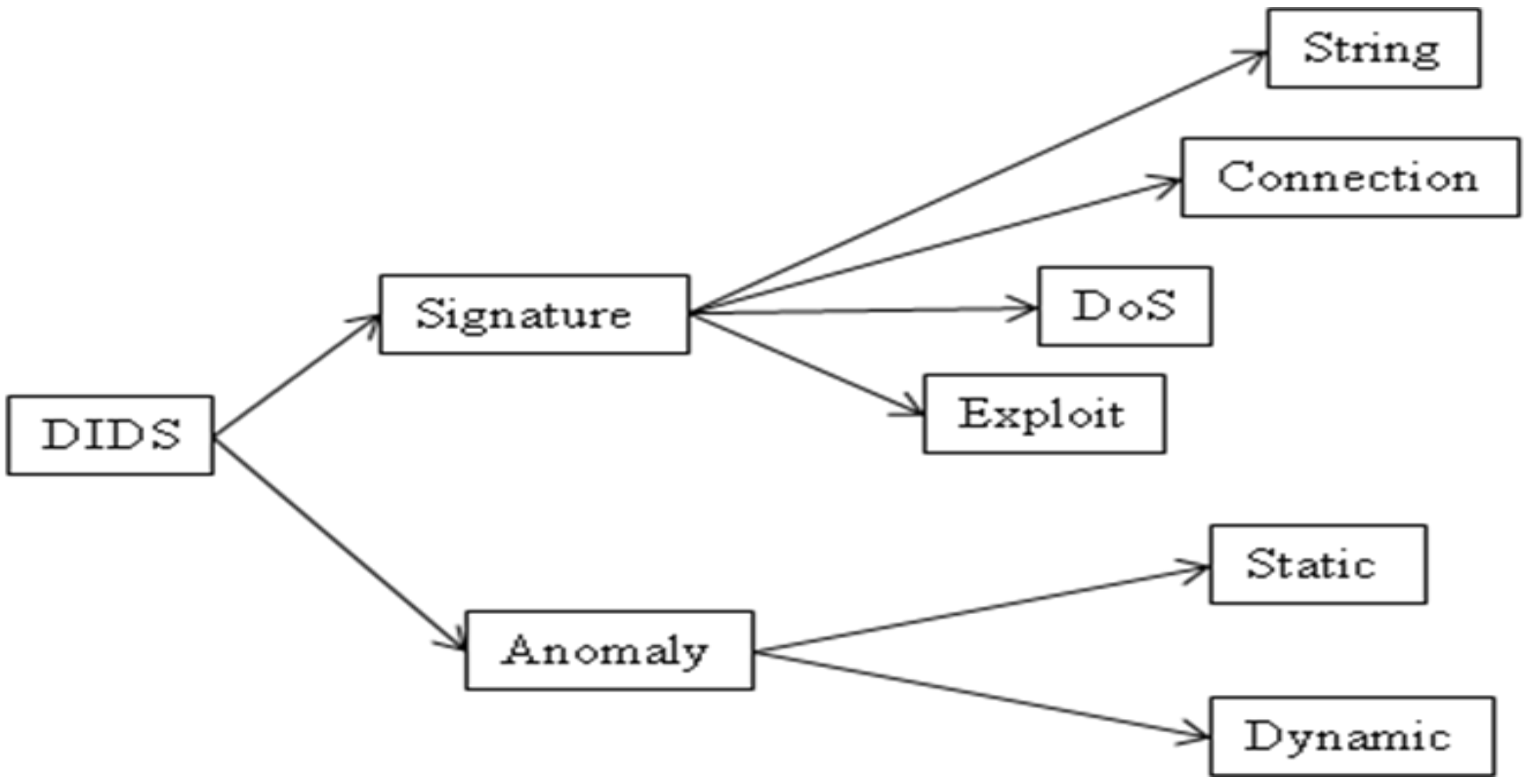

Fig 2.2. Mechanisms for intrusion detection

The work done in can be consulted for further reading [23] as it compared the various approaches using features, advantages, and disadvantages of each approach.

## 2.5. Review of Related Literatures

[5] proposed "an efficient hybrid system for anomaly detection in social networks". The model cascaded several machine learning algorithms that included decision tree, Support Vector Machine (SVM) and Naïve Bayesian classifier (NBC) for classifying normal and abnormal users on social networks. the anomaly detection engine uses SVM algorithm to classify social media network user as happy or disappointed, NBC algorithm is used based on a defined dictionary to classify social media users with social tendency. Unique features derived from users' profile and contents were extracted and used for training and testing of the model, performance evaluation conducted by experiment on the model using synthetic and real datasets from social network shows 98% accuracy.

(Mansoori et al. 2020) proposed "Suspicious Activity Detection of Twitter and Facebook using Sentimental Analysis". The model was designed using Natural

[24] proposed "Social Media Cyberbullying Detection using Machine Learning" the model leverage on the machine learning capability to detect the language patterns of bullies on social media network which will be used to generate a model that can automatically detect cyberbullying actions on the network. The model was evaluated using two classifier algorithms: SVM and Neural Network, TFIDF and sentiment analysis algorithms were used for features extraction. Different n-gram language models were used to evaluate the model and was found to achieved 92.8% accuracy using Neural Network with 3-grams and 90.3% accuracy using SVM with 4- grams while using both TFIDF and sentiment analysis together. NN performed better than the SVM classifier with average f-score of 91.9% while that of SVM achieved average f-score 89.8%.

(Shah, Sharma, and Bandgar 2021) proposed "Cybercrime Prevention on Social Media" in which a Facebook-like social network platform was designed as a proof-of-concept to detect cyberbullying and anormaly detection. Each user's post will be passed through "Post Classifier" to divide illegal posts from adult image post, the illegal post detection algorithm was used to check for violence-based objects. Similarly, sentiment analysis was used to check the intention of the message to detect any harrasment words in the content of the post. A threashold was set for any of the negative words and images posted by a user, if any illegal post by a user exceed a threashold, the user will be warned, and if there is persistence in the post, the user will be banned from making a post on the platform. Convolution Neural Network (CNN) machine learning algorithm was used in the implementation

[4] “Intrusion Detection System for Data Warehouse with Second Level Authentication”. This proposal was premised on the ground that earlier security mechanisms for data warehouse like “Role Based Access Controls (RBAC), Extended RBAC, Temporal RBAC (TRBAC), Risk-based access control, Intrusion Detection System (IDS) and some other customized security solutions for DWs does not detect a hacker that gain rightful access to the DW through credential theft. Therefore, a second level authentication mechanism within the IDS was proposed where a minute deviation from the user’s past behavior will be detected based on providing answer to secrete question that was provided at the account setup phase.

In their research work, [12] envisioned the most important criteria of employing social network in higher education, they observed broad view of possibilities for using social networks in higher education. The conceptual framework for academic social network should have the following four main objectives: (1) To provide academic service support and academic information dissemination; (2) To enable student support and communication; (3) To exalt social and cooperate learning; (4) To provide achievement representationability.

[13] developed an “Intrusion Detection System for College (ERP) Enterprise Resource Planning System” The system used a layered approach combined with a decision tree based architecture to detect attacks, the design and simulation was carried out using Netbeans integrated development environment with MYSQL database.

[8]proposed a model for building the network intrusion detection system using a machine learning algorithm called decision tree to detect anomaly based intrusion. The system used Recursive-Feature-Elimination (RFE) to select the best features from Change Control Intrusion Detection (CCIDS) 2017dataset. The dataset was split into training and test dataset, the training dataset was fit to the classification model developed to classify the test data as malicious or benign. The precision of the proposed system indicated True-Positive-Rate (TPR) of 99.9% and the False-Positive-Rate (FPR) of 0.1%

[14]proposed a dynamic Intelligent Intrusion Detection Model based on specific AI approach for intrusion detection. It is a hybrid system that combines anomaly, misuse and host based detection. SNORT packet sniffer was used for new data collection which will be passed through the inference engine to classify the operation as host-based or anomaly. The implementation algorithm used was neural networks and fuzzy logic with network profiling. Simple data mining techniques was used to process the network data to predict anomaly detection.

[7] proposed “Network Intrusion Detection System”. The system like any other intrusion detection system comprises of four major components which are: Traffic Collector for gathering activity and event data for analysis; Analysis Engine that analyzes the data that the traffic collector gathered; Signature Database which is an amalgamation of signatures known to be associated with suspicious and malicious activities; Management and Reporting Interface used by system administrators to manage the system and receive alerts when intrusions are detected. The system was implemented using Java programming language, used to detect specific attacks which are Man-in-the- Middle, DOS and ping of death.

[6] proposed “Data warehousing and data mining techniques for intrusion detection systems” the aim was to improve the performance and usability of Intrusion Detection Systems (IDS). The system model network traffic and alerts using a multi-dimensional data model and star schemas to perform network security analysis and to detect denial of service attacks. A prototype of the system was successfully tested at Army Research Labs.

[9] proposed “DBMS Application Layer Intrusion Detection for Data Warehouses (DW)”. They argued that the current DIDS lack capacity to detect heterogeneous oriented DW. In the research work, specific requirements for data warehouse based IDS was defined which lead to the proposal of a conceptual approach for a real-time DIDS for DWs at the SQL command level that works transparently as an extension of the Database Management System (DBMS) between the user applications and the database server itself.

In their review, [25] looked into four different approaches to intrusion detection in a network environment. The approaches are: Artificial Neural Network (ANN) “is comprised of a collection of processing elements that are highly interconnected, and convert a set of inputs to a set of desired outputs”; Self Organizing Map (SOM), Fuzzy Logic and Support Vector Machine (SVM).

Table 1. shows the summary of related literatures

| **Title [Ref]** | **Methodology/ Tools** | **Contribution** | **Research Gap** |
|---|---|---|---|
| An efficient hybrid system for anomaly detection in social networks [5] | R language and Python machine learning packages. | DT-SVMNB that classifies users as depressed one or suicidal one in the social network | Focus was on predicting vulnerable users on the social media not hackers |
| Suspicious Activity Detection of Twitter and Facebook using Sentimental Analysis [26] | Sentiment analysis, NLP based Part-Of-Speech (POS) tagging | Developed a model that can analyze the opinions posted on the internet to classify them as good, bad, or neutral. | Does not cover the hackers’ intrusion detection on the social media platforms |
| Social Media Cyberbullying Detection using Machine Learning [24] | SVM, Neural Network, TFIDF and sentiment analysis algorithms | Achieved average f-score of 91.9% and 89.8% for NN and SVM respectively on the detection. | Handled only one aspect of social media attacks – cyberbullying. Does not cover anomaly intrusion |
| Cybercrime Prevention on Social Media [27] | Django, NLP, NN, CNN | Developed a customized platform and implemented anomaly post detection | The intrusion detection can only detect cyberbullying attack |
| DBMS Application Layer Intrusion Detection for Data Warehouses [9] | Oracle 11g DBMS, TPC-H benchmark | Proposed DBMS Application Layer Intrusion Detection for Data Warehouses | Hackers with persistent attack can evade the security mechanism |
| Intrusion Detection System : Overview [25] | Qualitative | highlighted four different approaches to intrusion detection in a network environment | There was no proposed design |
| Intrusion detection system for data | MariaDB, MYSQL, TPC-H | Developed IDS for data warehouse with | Resilient hacker can use their hacking |
| warehouse with second level authentication [4] | | second level authentication to reinforce access control security | techniques to overcome the second level authentication |
| Intrusion Detection System for College ERP System [13] | JDK 1.7, MYSQL, NETBEANS | Developed IDS for college ERP system | The IDS only detect inconsistent data entry to the ERP system |
| Network Intrusion Detection System Using Machine Learning [8] | Python libraries, CCIDS, RFE | Achieved TPR of99.9% and FPR of 0.1% on CCIDS | Decision trees have a tendency to over-fit and can create biasness |
| Artificial Intelligence Techniques Applied To Intrusion Detection [14] | Data mining MySQL, SNORT, Fuzzy Logic | Used data mining techniques to process the network data used to predict anomaly detection. | Rule-based models are limited by the knowledge of the expert that developed sit |
| Network Intrusion Detection System [7] | Java, Traffic Sniffer | used to detect specific attacks which are Man-in-the- Middle, DOS and ping of death | Focused on NID for Man-in-the- Middle, DOS and ping of death, detection algorithm not specified |
| Data warehousing and data mining techniques for intrusion detection systems [6] | STAR schemas, OLAP | improved the performance and usability of Intrusion Detection Systems (IDS with the star schema | The improvement was for network intrusion, not DW intrusion detection |
| A Conceptual Model of Social Networking in Higher Education [12] | conceptual model | Identified 4 main functions of social network usage in higher education: academic service support; student support; social and cooperate learning; and achievement representation | Social vulnerability of the network was not explored. This will affect efficient learning |
| A Data Mining Approach for Attribute Selection in Intrusion Detection System [28] | WEKA | Justified that attribute selection in will improve Intrusion Detection System | IDS was not developed |

### 2.6. Research/Knowledge Gap

There is presently no developed intrusion detection system for social media platform to curtail the activities of hackers that have turned their attention to the platform. Most of the literatures reviewed do not have the intelligence to detect anomaly usage of social media account.

Role Based Access Control (RBAC), Extended RBAC, Temporal RBAC (TRBAC), Risk-based access control, etc does not have the ability to detect an attacker who obtains access to the system using some compromised credentials. Intrusion Detection System (IDS) and some other customized security solutions for DWs have also been proposed including second level authentication. But the same mechanism used to evade the first level authentication can still be employed to overcome the second level authentication security approach. Hence fooling the attacker with fake response will provide a better reinforcement to DW intrusion detection/prevention mechanism.

Most of the previous proposals used KDD-CUP-99and DARPA 98/99 dataset for training but these datasets have become outdated with limitations over updating of new attacks [8]. Those earlier models might not work well owing to the fact that attackers changes their signatures regularly to evade detection.

## 3. CONCLUSIONS

Social media network has become the nerve centre of the virtual community that connects billions of heterogeneous users for mutual interaction. Because of its dynamic nature where users can share contents freely between friends and followers, hackers are seriously exploiting this rich platform for malicious intention. Various strategies for attacking social media users have been highlighted in this paper with various preventive approaches proposed by researchers. Despite all these preventive approaches, hackers activities on the platform is on the rise hence social media intrusion detection system will be highly recommended as second line of defence against the attacks of hackers on the social media network platforms.

## ACKNOWLEDGEMENTS

The authors would like to thank my lecturers for their constructive criticism and impact!

## AUTHORS

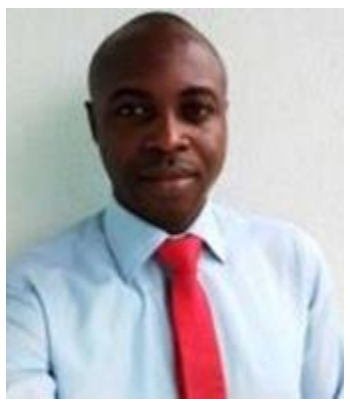

**Emmanuel Etuh** is currently pursuing a PhD in Computer Science at the University of Nigeria, Nsukka, He obtained his first degree certificate in Computer Science from Kogi State University, Anyigba in 2009 and an MSc degree in Computer Science from Ahmadu Bello University, Zaria in 2014, His research interest include Artificial Intelligence, CyberSecurity, and Software Engineering.

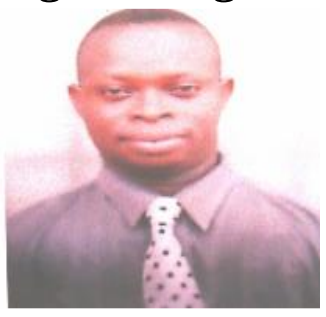

**Professor Bakpo**, Francis Sunday received his M Eng. degree in Computer Science and Engineering from Kazakh National Technical University, Almaty (formerly USSR) in 1994 and PhD degree in Computer Engineering in 2008 from Enugu State University of Science and Technology (ESUT), Agbani. He joined the Department of Computer Science at University of Nigeria Nsukka as lecturer II in June 1996 and further progressed from the rank of lecturer II to Professor in 2010. His current research interest includes Computer architectures, computer communications networking, artificial neural network applications, and Petri net theory. He is dully registered professionally as member Nigeria Computer Society, MNCS and Computer professional of Nigeria, MCPN, respectively and has also published a number of excellent journal papers, books and conference proceeding papers in his field.

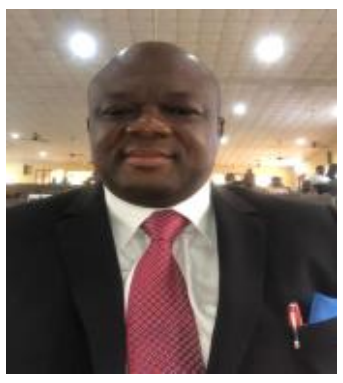

**Eneh, Agozie H** is a Senior Lecturer in the Department of Computer Science at the University of Nigeria, Nsukka. His Area of Specialization include: authentication protocols analysis, network security, optimisation theories, medical informatics, and assistive technologies for educating children and adolescents with learning difficulties.